\documentstyle[12pt]{article}
\newcommand{\nc}{\newcommand}
\nc{\postscript}[2]
{\setlength{\epsfxsize}{#2\hsize}\centerline{\epsfbox{#1}}}
\nc{\bg}{B. Grzadkowski}
\nc{\non}{\nonumber}
\def\dps{\displaystyle}
\def\mib#1{\mbox{\boldmath $#1$}}

\def\bra#1{\langle #1 |} \def\ket#1{|#1\rangle}
\def\vev#1{\langle #1\rangle}

\nc{\barx}{\bar{x}}\nc{\pbarn}{\;\hbox {pb}}\nc{\fbarn}{\;\hbox {fb}}
\nc{\hc}{\hbox {h.c.}} \nc{\re}{\hbox {Re}} 
\nc{\mev}{\hbox {MeV}} \nc{\gev}{\;\hbox {GeV}}
\def\gesim{\lower0.5ex\hbox{$\:\buildrel >\over\sim\:$}}
\def\lesim{\lower0.5ex\hbox{$\:\buildrel <\over\sim\:$}}
\nc{\prd}[3]{{\it Phys.\ Rev.}\ {{\bf D{#1}} (#2), #3}}
\nc{\prl}[3]{{\it Phys.\ Rev.\ Lett.}\ {{\bf {#1}} (#2), #3}}
\nc{\plb}[3]{{\it Phys.\ Lett.}\ {{\bf B{#1}} (#2), #3}}
\nc{\npb}[3]{{\it Nucl.\ Phys.}\ {{\bf B{#1}} (#2), #3}}
\nc{\ptp}[3]{{\it Prog.\ Theor.\ Phys.}\ {{\bf {#1}} (#2), #3}}
\nc{\zfp}[3]{{\it Z.\ Phys.}\ {{\bf C{#1}} (#2), #3}}
\nc{\epj}[3]{{\it Eur.\ Phys.\ J.}\ {{\bf C{#1}} (#2), #3}}
\nc{\mpla}[3]{{\it Mod.\ Phys.\ Lett.}\ {{\bf A{#1}} (#2), #3}}
\nc{\rmp}[3]{{\it Rev.\ Mod.\ Phys.}\ {{\bf {#1}} (#2), #3}}
\nc{\ijmpa}[3]{{\it Int.\ J.\ of\ Mod.\ Phys.}\
               {{\bf A{#1}} (#2), #3}}
\nc{\ttbar}{t\bar{t}}         \nc{\bbbar}{b\bar{b}}
\nc{\tanb}{\tan \beta}        \nc{\twbdec}{t\to W^+ b}
\nc{\tbwbdec}{\bar{t}\to W^- \bar{b}}
\nc{\epem}{e^+e^-}            \nc{\eett}{\epem \to \ttbar}
\nc{\sigeett}{\sigma_{e\bar{e}\to\ttbar}}
\nc{\wpwm}{W^+W^-}            \nc{\tbar}{\bar{t}}
\nc{\bbar}{\bar{b}}           \nc{\wpp}{W^+}
\nc{\mt}{m_t}    \nc{\mts}{m_t^2}   \nc{\mw}{m_W}    \nc{\mws}{m_W^2}
\nc{\mz}{m_Z}    \nc{\mzs}{m_Z^2}
\nc{\ttbardec}{\ttbar \to W^+W^-\bbbar}
\nc{\wwbb}{W^+W^-\bbbar}      \nc{\sm}{SM}
\nc{\cw}{\cos\theta_W}        \nc{\sw}{\sin\theta_W}
\nc{\sws}{\sin^2\theta_W}     \nc{\sig}{\sigma_{tot}}
\nc{\lp}{{\ell}^+}              \nc{\lm}{{\ell}^-}
\nc{\epsl}{\epsilon_L}        \nc{\cp}{C\!P}
\nc{\splus}{s_+}       \nc{\smin}{s_-}        \nc{\eps}{\epsilon}
\nc{\psp}{Ps_+}        \nc{\psm}{Ps_-}        \nc{\lsp}{ls_+}
\nc{\lsm}{ls_-}        \nc{\sss}{s_+s_-}      \nc{\m}{m_t}
\nc{\mq}{m_t^2}        \nc{\mr}{\frac{1}{\m}} \nc{\av}{A_{\gamma}}
\nc{\bv}{B_{\gamma}}   \nc{\az}{A_Z}          \nc{\bz}{B_Z}
\nc{\avs}{A_{\gamma}^2}\nc{\azs}{A_Z^2}       \nc{\bzs}{B_Z^2}
\nc{\dav}{\delta \! A_{\gamma}}   \nc{\dbv}{\delta \! B_{\gamma}}
\nc{\dcv}{\delta C_{\gamma}}      \nc{\ddv}{\delta \! D_{\gamma}}
\nc{\daz}{\delta \! A_Z}          \nc{\dbz}{\delta \! B_Z}
\nc{\dcz}{\delta C_Z}             \nc{\ddz}{\delta \! D_Z}
\nc{\dev}{\delta \! E_{\gamma}}   \nc{\dez}{\delta \! E_Z}
\nc{\dfv}{\delta \! F_{\gamma}}   \nc{\dfz}{\delta \! F_Z}
\nc{\rdav}{{\rm Re}(\delta \! A_{\gamma}) \:}
\nc{\rdbv}{{\rm Re}(\delta \! B_{\gamma}) \:}
\nc{\rdcv}{{\rm Re}(\delta C_{\gamma}) \:}
\nc{\rddv}{{\rm Re}(\delta \! D_{\gamma}) \:}
\nc{\rdaz}{{\rm Re}(\delta \! A_Z) \:}
\nc{\rdbz}{{\rm Re}(\delta \! B_Z) \:}
\nc{\rdcz}{{\rm Re}(\delta C_Z) \:}
\nc{\rddz}{{\rm Re}(\delta \! D_Z) \:}
\nc{\idav}{{\rm Im}(\delta \! A_{\gamma}) \:}
\nc{\idbv}{{\rm Im}(\delta \! B_{\gamma}) \:}
\nc{\idcv}{{\rm Im}(\delta C_{\gamma}) \:}
\nc{\iddv}{{\rm Im}(\delta \! D_{\gamma}) \:}
\nc{\idaz}{{\rm Im}(\delta \! A_Z) \:}
\nc{\idbz}{{\rm Im}(\delta \! B_Z) \:}
\nc{\idcz}{{\rm Im}(\delta C_Z) \:}
\nc{\iddz}{{\rm Im}(\delta \! D_Z) \:}
\nc{\cz}{(1+v_e^2)d\:\!'^2}         \nc{\ci}{v_ed\:\!'}
\nc{\ccz}{v_ed\:\!'^2}              \nc{\cci}{d\:\!'}
\nc{\lspace}{\;\;\;\;\;\;\;\;\;\;}  \nc{\llspace}{\lspace \lspace}
\nc{\beq}{\begin{equation}}   \nc{\eeq}{\end{equation}}
\nc{\bea}{\begin{eqnarray}}   \nc{\eea}{\end{eqnarray}}
\nc{\baa}{\begin{array}}      \nc{\eaa}{\end{array}}
\nc{\bit}{\begin{itemize}}    \nc{\eit}{\end{itemize}}
\nc{\ben}{\begin{enumerate}}  \nc{\een}{\end{enumerate}}
\nc{\ocal}{{\cal O}}
\begin{document}
\pagestyle{empty} \setlength{\footskip}{2.0cm}
\setlength{\oddsidemargin}{0.5cm} \setlength{\evensidemargin}{0.5cm}
\renewcommand{\thepage}{-- \arabic{page} --}
\def\mib#1{\mbox{\boldmath $#1$}}
\def\bra#1{\langle #1 |}      \def\ket#1{|#1\rangle}
\def\vev#1{\langle #1\rangle} \def\dps{\displaystyle}
\nc{\tb}{\stackrel{{\scriptscriptstyle (-)}}{t}}
\nc{\bb}{\stackrel{{\scriptscriptstyle (-)}}{b}}
\nc{\fb}{\stackrel{{\scriptscriptstyle (-)}}{f}}
\nc{\pp}{\gamma \gamma}
\nc{\pptt}{\pp \to \ttbar}
   \def\thebibliography#1{\centerline{REFERENCES}
     \list{[\arabic{enumi}]}{\settowidth\labelwidth{[#1]}\leftmargin
     \labelwidth\advance\leftmargin\labelsep\usecounter{enumi}}
     \def\newblock{\hskip .11em plus .33em minus -.07em}\sloppy
     \clubpenalty4000\widowpenalty4000\sfcode`\.=1000\relax}\let
     \endthebibliography=\endlist
   \def\sec#1{\addtocounter{section}{1}\section*{\hspace*{-0.72cm}
     \normalsize\bf\arabic{section}.$\;$#1}\vspace*{-0.3cm}}
\vspace*{-0.7cm}
\begin{flushright}
$\vcenter{
\hbox{TOKUSHIMA Report}
\hbox{(hep-ph/0406247)}
}$
\end{flushright}

\vskip 1.7cm
\renewcommand{\thefootnote}{*}

\begin{center}
{\large\bf New-Physics Search in $\mib{\gamma\gamma\to t\bar{t}}$}
\footnote{Talk at {\it International Workshop on Physics and
 Experiments with Future Electron-Positron Linear Colliders
 (LCWS2004)}, April 19-23, 2004, Le Carre des Sciences, Paris,
 France. \\
This work is based on collaboration with
B. Grzadkowski (Warsaw U), K.Ohkuma (Fukui U Tech) and
J. Wudka (UC Riverside).}
\end{center}

\vspace*{1cm}
\renewcommand{\thefootnote}{*)}
\begin{center}
{\sc Zenr\=o HIOKI$^{\:}$}\footnote{E-mail address:
\tt hioki@ias.tokushima-u.ac.jp}
\end{center}

\vspace*{0.7cm}
\centerline{\sl Institute of Theoretical Physics,\ 
University of Tokushima}

\vskip 0.14cm
\centerline{\sl Tokushima 770-8502, JAPAN}

\vspace*{2.9cm}
\centerline{ABSTRACT}

\vspace*{0.6cm}
\baselineskip=20pt plus 0.1pt minus 0.1pt
We performed an analysis on possible anomalous top-quark
couplings generated by $SU(2)\times U(1)$ gauge-in\-var\-i\-ant
dimension-6 effective operators, applying the optimal-observable
procedure to the final lepton/$b$-quark momentum distribution in
$\gamma\gamma\to t\tbar \to \ell X/bX$. We studied how many
such anomalous coupling constants could be determined
altogether through these distributions.

\vspace*{0.4cm} \vfill

\newpage
\renewcommand{\thefootnote}{\sharp\arabic{footnote}}
\pagestyle{plain} \setcounter{footnote}{0}
\baselineskip=21.0pt plus 0.2pt minus 0.1pt
Linear colliders of $\epem$ are expected to work as top-quark
factories, and therefore a lot of attention has been paid to
study possible non-standard top-quark interactions in
$e\bar{e}\to t \bar{t}$.
An interesting option for such $\epem$ machines would be that
of $\gamma\gamma$ collisions, where initial energetic photons
are produced via electron and laser-light backward
scatterings [1].

Indeed a number of authors have already considered top-quark
production and decays in $\gamma\gamma$ collisions in order to
study {\it i}) Higgs-boson couplings to the top quark and
photon, or
{\it ii}) anomalous top-quark couplings to the photon
(see the ref.list of [2] on relevant papers throughout this
report). However, what we will observe in real
experiments are combined signals that originate both from the
process of top-quark production and, {\it in addition}, from
its decays. Therefore, in our latest articles [2]
we studied $\gamma\gamma \to \ttbar \to {\ell} X/bX$, including
all possible non-standard interactions together (production and
decay), and performed a comprehensive analysis as
model-independently as possible. I will give a brief sketch of
this work here.

In order to describe possible new-physics effects, we used
an effective low-energy Lagrangian [3], 
i.e., the Standard-Model (SM) Lagrangian modified by the
addition of a
series of $SU(2)\times U(1)$ gauge-invariant operators
${\cal O}_i$, which are suppressed by inverse powers of a
new-physics scale ${\mit\Lambda}$ and whose coefficients
$\alpha_i$ parameterize the low-energy
effects of this new physics.

Among these ${\cal O}_i$, the largest contribution comes from
dimension-6 operators, and they are classified into those producing
(1) $C\!P$-conserving $t\bar{t}\gamma$ vertex,
(2) $C\!P$-violating $t\bar{t}\gamma$ vertex,
(3) $C\!P$-conserving $\gamma\gamma H$ vertex,
(4) $C\!P$-violating $\gamma\gamma H$ vertex,
and
(5) anomalous $tbW$ vertex.
On the other hand, the $\nu\ell W$ vertex is assumed to be
properly described within the SM.

In order to derive distributions of secondary fermions we have
applied the Kawasaki--Shirafuji--Tsai formula [4] 
with FORM [5]
for algebraic manipulations, treating the decaying $t$
and $W$ as on-shell particles. We neglected contributions that are
quadratic in non-standard interactions, therefore the angular-energy
distributions of the $\ell/b$ in the $e\bar{e}$ CM frame can be
expressed as
\begin{equation}
\frac{d\sigma}{dE_{\ell/b} d\cos\theta_{\ell/b}}
=f_{\rm SM}(E_{\ell/b}, \cos\theta_{\ell/b})
 + \sum_i \alpha_i f_i (E_{\ell/b}, \cos\theta_{\ell/b}),
\label{distribution}
\end{equation}
where $f_{\rm SM}$ and $f_i$ ($i=\gamma 1,\gamma 2, h1,h2,d$) are
calculable functions: $f_{\rm SM}$ denotes the standard-model
contribution, $f_{\gamma 1,\gamma 2}$ describe the anomalous
$C\!P$-conserving and $C\!P$-violating
$t\bar{t}\gamma$-ver\-ti\-ces contributions respectively,
$f_{h1,h2}$  those generated by the anomalous $C\!P$-conserving
and $C\!P$-violating $\gamma\gamma H$-ver\-ti\-ces,
and $f_d$ that by the anomalous $tbW$-vertex.
The initial-state polarizations are characterized
by the incident electron and positron longitudinal polarizations
$P_e$ and $P_{\bar{e}}$, the average
helicities of the initial-laser photons $P_{\gamma}$ and
$P_{\tilde{\gamma}}$, and their maximum average linear polarizations
$P_t$ and $P_{\tilde{t}}$.

We applied the optimal-observable procedure [6]
to eq.(\ref{distribution})
assuming $\sqrt{s_{e\bar{e}}}=500$ GeV,
and tried to estimate the expected statistical precision for
$\alpha_i$ determination for both
linear and circular $e/\gamma$ polarizations. Leaving its detail
to our article [2], 
let me here briefly summarize the
result:
We found no numerically-stable solutions in analyses with
three (or more) free parameters although we first aimed to
determine the five parameters $\alpha_i$
altogether.\footnote{Note however that our result never means
    there remains no longer any hope for such analyses with three
    or more free parameters since we only used
    some typical beam polarizations. We should study various
    other beam polarizations comprehensively to get the final
    answer.}\ 
We however found stable solutions in two-parameter analyses
for the following sets:
($\alpha_{\gamma 2}$, $\alpha_d$), ($\alpha_{h 1}$, $\alpha_d$),
($\alpha_{h 2}$, $\alpha_d$) for $\ell X$, and
($\alpha_{\gamma 2}$, $\alpha_{h1}$), ($\alpha_{\gamma 2}$, $\alpha_d$),
($\alpha_{h1}$, $\alpha_{h2}$), ($\alpha_{h1}$, $\alpha_d$), 
($\alpha_{h2}$, $\alpha_d$) for $bX$.

We did not find any
solution that would allow for a determination of $\alpha_{\gamma 1}$
for either $\ell X$ or $bX$. We therefore have to look
for other suitable processes to determine this parameter.
The precision of $\alpha_{\gamma 2}$ is not very good
either, but using $b$-quark gives still much better result
than in the case of the
lepton analysis. On the other hand, we found that analyzing
the $b$-quark process with linearly polarized beams enables
us to estimate some $\alpha_i$ that were
unstable in the lepton analysis, i.e.,
cases ($\alpha_{\gamma 2}$, $\alpha_{h1}$) and ($\alpha_{h1}$, $\alpha_{h2}$).
One of them, ($\alpha_{h1}$, $\alpha_{h2}$), is especially useful to probe the
$C\!P$ properties of heavy Higgs bosons.
As for the determination of $\alpha_d$, the ${\ell}X$ final state
seems to be more appropriate. These comparisons
show that both final states ($bX$ and ${\ell}X$)
provide complementary information and should therefore be
included in a complete analysis.

Some non-standard couplings, which should be determined here,
could also be studied in the standard $e^+e^-$ option of a
linear collider. Therefore, it is worth while to compare the potential
power of the two options. As far as the parameter $\alpha_{\gamma 1}$
is concerned, the $\gamma\gamma$ collider does not allow for its
determination, while it could be determined at $e^+e^-$.
The second $\ttbar\gamma$ coupling $\alpha_{\gamma 2}$,
which is proportional to the real part of the
top-quark electric dipole moment,
can be measured here. 
For the measurement of $\gamma\gamma H$
couplings, $e^+e^-$ colliders are, of course, useless, while here,
for the $bX$ final state both $\alpha_{h 1}$ and $\alpha_{h 2}$
could be measured. In the case of the decay form factor $\alpha_d$
measurement, the $e^+e^-$ option seems to be a little more advantageous,
especially if $e^+e^-$ polarization can be tuned
appropriately, which we found in our previous work.

It should be emphasized finally that our effective-operator
strategy is valid only for ${\mit\Lambda}
\gg v\simeq 250 \gev$, in contrast to the analysis of
$\epem \to t\bar{t}\to {\ell}X$.
Should the reaction $\gamma \gamma \to \ttbar \to \ell X/b X$
exhibit a deviation from the SM predictions  that cannot be
described properly within this framework, we would have an
indication of low-energy beyond-the-SM physics, e.g.,
two-Higgs-doublet models with new
scalar degrees of freedom of relatively low mass scale.

\vskip 1cm
 \def\NCA{\em Nuovo Cimento}
 \def\NIM{\em Nucl. Instrum. Methods}
 \def\NIMA{{\em Nucl. Instrum. Methods} {\bf A}}
 \def\NPB{{\em Nucl. Phys.} {\bf B}}
 \def\PLB{{\em Phys. Lett.} {\bf B}}
 \def\PRL{\em Phys. Rev. Lett.}
 \def\PRD{{\em Phys. Rev.} {\bf D}}
 \def\EPJ{{\em Eur. Phys. J.} {\bf C}}
 \def\ZP{{\em Z. Phys.} {\bf C}}
%


\begin{thebibliography}{99}
\bibitem{Ginzburg:1981vm}
I.F. Ginzburg, G.L. Kotkin, V.G. Serbo and V.I. Telnov,
{\em Nucl. Instrum. Meth.}  {\bf 205} (1983) 47.
%
\bibitem{GHOW}{B.~Grzadkowski, Z.~Hioki, K.~Ohkuma and J.~Wudka,
\NPB \rm {\bf 689} (2004) 108 (hep-ph/0310159)};
{\PLB\ \rm {\bf 593} (2004) 189 (hep-ph/0403174)}.
%
\bibitem{Buchmuller:1986jz}
{W.~Buchm\"uller and D.~Wyler, \NPB \rm {\bf 268} (1986) 621}.
%
\bibitem{technique}
{Y.S. Tsai, \PRD \rm {\bf 4} (1971) 2821};
{\it ibid}.\ {\bf 13} (1976) 771(Erratum);\\
{S. Kawasaki, T. Shirafuji and S.Y. Tsai,
{\em Prog. Theor. Phys.} \rm {\bf 49} (1973) 1656}.
%
\bibitem{FORM}
J.A.M. Vermaseren, ``{\it Symbolic Manipulation with FORM} ", version 2,
Tutorial and Reference Manual, CAN, Amsterdam 1991, ISBN 90-74116-01-9.
%
\bibitem{optimal}
{D.~Atwood and A.~Soni, \PRD \rm {\bf 45} (1992) 2405}; \\
{M.~Davier, L.~Duflot, F.~Le Diberder and A.~Rouge,
\PLB \rm {\bf 306} (1993) 411}; \\
{M.~Diehl and O.~Nachtmann, \ZP \rm {62} (1994) 397}; \\
{J.F.~Gunion, B.~Grzadkowski and X.G.~He,
\PRL\ \rm {\bf 77} (1996) 5172} (hep-ph/9605326).%
%
\end{thebibliography}
\end{document}